\documentclass[prl,twocolumn,showpacs,showkeys,preprintnumbers,amsmath,amssymb]{revtex4}

 \usepackage{epsf}

\usepackage{amsmath,amsfonts}

\newcommand{\tr}{\hbox{ Tr}}

\begin{document}

\title{The Minimal Quiver Standard Model}
\author{David Berenstein$^{1}$}
\email{dberens@physics.ucsb.edu}
\author{Samuel Pinansky$^1$}
\email{samuelp@physics.ucsb.edu}

 \affiliation{$^1$Department of
Physics, University of California
at Santa Barbara, CA 93106}

\begin{abstract}
This paper discusses the minimal quiver gauge theory embedding of the standard model that could arise from brane world type string theory constructions. It is based on the low energy effective field theory of D-branes in the perturbative regime. The model
differs from the standard model by the addition of one extra massive gauge boson, 
and contains only one additional parameter to the standard model: the mass of this new particle. The coupling  of this new particle to the standard model is uniquely determined by input from the standard model and consistency conditions of perturbative string theory. We also study some aspects of the phenomenology of this model and bounds on its possible
observation at the Large Hadron Collider. 
\end{abstract}
 \maketitle

String Theory has long had a goal of unifying the Standard Model of particle physics with gravity in a consistent way. 
The possibility of a huge number of vacua where, in principle, anything can happen, suggests that we will find where we are in this landscape of vacua by doing sufficiently many experiments to determine the shape of the extra dimensions.
If string theory is to be verified experimentally, it is useful to ask whether the Large Hadron Collider experiment will be able to see stringy physics directly and how to differentiate such effects from other results. Here we are limited by our 
calculational abilities. 

For these purposes, the most interesting scenarios for accelerator experiments are those where the string theory is perturbative, and the string scale is relatively low (a few TeV). 
The vast majority of these
models are based on D-branes. 
Thus, we will ask a more limited question: Do low string scale D-brane models have universal low energy predictions for physics beyond the Standard Model that one can use to decide whether or not we live on a D-brane?

The answer to this question is yes, but most of the tests rely on simple group theory properties of the spectrum, or directly observing the string theory resonances in an experiment. We want to find a uniquely stringy prediction that does not follow just from low energy effective field theory with the required matter content.

The most important observation towards this goal is that the gauge group on a stack of identical D-branes is never $SU(N_i)$ for $N_i>2$, but  $U(N_i)$ \cite{Pol}. 

Therefore, at tree level, the coupling constant of the $U(1)$ part of the group and the $SU(N)$ part of the  $U(N)$ gauge group are related, as well as the particles' charge under the $U(1)$ and $SU(N)$, respectively. This is not required in pure effective field theory and constitutes a prediction for particle physics beyond the standard model based on string theory ideas. Our purpose is to explore this idea in the simplest possible setup compatible within an abstract D-brane setup.

We will show that the minimal model compatible with perturbative D-brane string dynamics that contains the standard model
has only one extra particle beyond the standard model. It is a massive vector boson that extends the color gauge group $SU(3)_c$ to $U(3)$ and the extra $U(1)$ acts essentially as a gauged baryon number symmetry. This extension is required for any D-brane model. What we find is a unique model that requires just that one extra particle and, moreover, all the couplings of this particle to matter are calculable from standard model parameters which are fixed by experimental observation. We will call this model the Minimal Quiver Standard Model (MQSM).
The model we find is very similar to models studied in \cite{Kir,Cremades:2003qj,CIK}, but it is simpler and therefore more predictive.  For a review of D-brane models, see \cite{BCLS}.

\begin{center}{\em The Model}
\end{center}

We will imagine starting with an abstract model of point-like branes at singularities, in the spirit of \cite{AIQU,BJL}. In this way we avoid the geometric Kaluza-Klein tower of states from extended branes. Also, we will not address the issues of global consistency and tadpole cancellation.  Instead, we will try to work in a model independent way and use consistency conditions
of the couplings to do effective field theory for physics beyond the standard model, similar to \cite{AKT}. 

The constraints we are setting also allow for a low effective string scale in the vicinity of the standard model singularity by warping the extra dimensions and solving some hierarchy problems a la Randall-Sundrum \cite{RS}. This allows us to avoid the problems older toroidal based compactifications have with the gauge couplings at a low string-scale.  In spirit, we are following abstract conformal field theory techniques for unoriented models in the $\alpha'\to 0$ limit, but with no supersymmetry. We will also require that the abstract model allow all of the  coupling constants of the standard model perturbatively. 

D-Brane constructions give a number of building block rules that we can use to build up the standard model or something very close to it.  
In  orientifold brane configurations, branes come in two different types.  There are the branes whose images under the orientifold are different from themselves and their image branes, and also branes who are their own images under the orientifold procedure.  Stacks of the first type combine with their mirrors and give rise to $U(N)$ gauge groups, while stacks of the second type give rise to only $SO(N)$ or $Sp(N)$ gauge groups.  All matter states arise as bifundamental representations (they have two indices)  $(N_a,\overline{N}_b)$, or $(N_a,N_b)$ and their complex conjugates where the stacks of branes intersect each other. 

The standard model group $SU(3)\times SU(2)\times U(1)\sim U(3)\times Sp(1)$ can be accommodated in a two-stack model. However, the matter content can not, as the leptons are doublets under $Sp(1)$ and are not charged under color.  This means we need to extend the gage group of the standard model. The minimal extension requires us to enlarge the gauge group by the smallest amount possible,
 giving a gauge group $U(3)\times Sp(1)\times U(1)$ that has three stacks of branes. The other possibility $U(3)\times U(2)$ is ruled out. That the $SU(2)$ weak group can be described as $Sp(1)$ in string models has been advocated in \cite{Cremades:2003qj}, as this reduces the required number of Higgs doublets to generate all Yukawa couplings at tree level.

\begin{figure}[htp]
\begin{center}
\epsfysize=4.0cm\epsffile{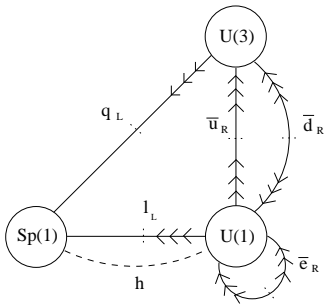} 
\parbox{8.9cm}{\caption{Quiver representation for the Minimal Quiver Standard Model. The arrow directions indicate fundamental or antifundamental representations for the $U(N)$ gauge groups}\label{fig:quiver}}
\end{center}
\end{figure}

Using this gauge group there is only one choice for the left handed quark doublet $q_L$, as a bifundamental $(3,2)_0$.  Since we know the right handed quarks have different hypercharge we make $\overline{u}_R$ transform as $({\bar 3},1)_1$ and $\overline{d}_R$ as $({\bar 3},1)_{-1}$. As described in \cite{Beren}, if we make the right handed quarks appear as two index representations of $SU(3)$ (like they do in $SU(5)$ GUT models), the Yukawa couplings for some of the quarks are forbidden.
Our choice also eliminates the cubic non-abelian anomaly for the $U(3)$ stack.  The lepton doublet $\ell_L$ is $(1,2)_{1}$, which leaves only the right handed electron to fit. The
$\overline{e}_R$ need to come from strings stretched between the $U(1)$ stack and its mirror, giving a field in a $(1,1)_{-2}$ representation (for a $U(N)$ stack, these are in the symmetric representation).  Finally, we can have a scalar Higgs field with the appropriate quantum numbers coming from strings stretching between the $Sp(1)$ and $U(1)$ stacks.  The quiver summarizing the spectrum is in fig. (\ref{fig:quiver}).

It is easy to check that anomalies factorize. However,  the string consistency conditions (RR tadpole cancellations) are stronger than anomaly cancelation. The modified constraints can be found in \cite{Kir}, and 
it is easy to verify that in this model such conditions are satisfied.

In the end, we have mixed anomalies between the two $U(1)$'s and the $SU(N)$ groups.  It is simple to show that the combination $Q_{U(3)}-3Q_{U(1)}$ has no mixed anomaly, and as expected, the hypercharge $Q_Y\equiv \frac{1}{6}Q_{U(3)}-\frac{1}{2}Q_{U(1)}$ is anomaly free.  However, the $Q_{U(3)}$ is not anomaly free, and we expect this anomaly to be canceled by a Green-Schwarz mechanism, where an extra coupling to a $RR$ two-form field or axion to the anomalous $U(1)$ will cancel the anomaly and give a mass to the $U(1)$.  This massive $U(1)$ gauge boson is the only extra particle beyond the standard model which this model predicts, and its mass is dependent on the specific geometry of the string construction. For typical models, the mass is of order $g^{1/2}_s M_s$, and it is smaller than the string scale (let us say, around $M_s/10$). Thus, it is justified to keep it in the low energy action and to ignore all massive string modes in the effective field theory.
Also, since the $U(1)$ is made massive by $RR$ couplings, the $U(1)$ symmetry remains as a global perturbative symmetry, which in this case is simply the baryon number.  Thus in this model the proton is perturbatively stable \cite{IQ}.

Looking at the quantum numbers we can write the most general Yukawa interaction as:
\begin{equation}
y_{ij}^uq_{L}^i\overline{u}^j_{R}h+y_{ij}^dq^i_{L}\overline{d}_{R}^jh^*+y_{ij}^e \ell^i_L\overline{e}_{R}^jh^*+{\rm h.c.}.
\end{equation}
This is the usual coupling of the Higgs particle to the standard model, and all Yukawa coupling constants are allowed at disk level. This makes our candidate model suitable for a perturbative D-brane extension of the standard model. The model we have just described is the MQSM.

To our knowledge, this model is simpler than all other models found with D-brane configurations so far \cite{AKT1,Kir,Ibanez:2001nd}.  Groups of type $USpU$ with the right chiral matter have been identified
in \cite{Kir}, but they also have extra non-chiral fermions and do not satisfy the tadpole conditions. Finding a D-brane configuration with the right matter content remains as an open problem.

Since we only have one Higgs doublet,  all of the Yukawa couplings are fully constrained by experimental data.
In our construction all couplings constants of the model, except for the mass
of the extra vector particle, are fixed by the standard model coupling constants. In this sense, 
our minimal model has only one free parameter and is very predictive.
 
\begin{center}\em Mass of the $Z'$
\end{center}

The MQSM model only contains a single extra neutral gauge boson with respect to the Standard model. We can calculate its mixing with the $Z$ and mass matrix exactly.  We label the gauge fields of the  $SU(3)\times U(1)\times SU(2)\times U(1)$,  as $C^a_\mu,C_\mu,A^a_\mu,B_\mu$ with couplings $g_3,g_3/\sqrt{6},g_2,g_1$ respectively.  In stringy conventions, the kinetic term for the $U(3)$ part splits up as:
\begin{equation}
-\frac{1}{2g_3^2}\tr({\bf C_{\mu\nu}C^{\mu\nu}})=-\frac{1}{4g_3^2}\sum_{a=1}^8C_{\mu\nu}^aC^{\mu\nu}{}^a-\frac{3}{2g_3^2}C_{\mu\nu}C^{\mu\nu}
\end{equation}
since we are using the standard normalization for the $SU(3)$ generators. The relations between the $SU(3)$ coupling and the $C_\mu$ coupling is dictated by the kinetic term being given by a single trace. This type of relation is string theoretic in nature and does not follow from
effective field theory. Because of RG flow effects, we set this relation at the string scale, where all the gauge groups are weakly coupled.

The kinetic terms for the $SU(2)\times U(1)$ are 
\begin{equation}
-\frac{1}{4g_2^2}\sum_{a=1}^3 F_{\mu\nu}^aF^{\mu\nu}{}^a-\frac{1}{4g_1^2}B_{\mu\nu}B^{\mu\nu}.
\end{equation}
The massless and massive direction of the $U(1)$'s  (before electroweak symmetry breaking) are given by
\begin{align}
Y_\mu&\equiv \sin\theta_P \hat C_\mu-\cos\theta_P \hat B_\mu,\quad Y'_\mu\equiv \cos\theta_P \hat C_\mu+\sin\theta_P \hat B_\mu.
\end{align}
where the $\hat C, \hat B$ are canonically normalized and  $\theta_P$ is
\begin{equation}
\theta_P\equiv \tan^{-1}\left(\sqrt{\frac{2}{3}}\frac{g_1}{g_3}\right)
\end{equation}
This comes from requiring that the massless set of gauge bosons are anomaly free, so the hypercharge gauge boson and its orthogonal complement are singled out. There is an explicit mass term in the lagrangian from the Green-Schwarz mechanism for the new gauge field $(-1/2) M ^2 Y'_\mu {Y'}^\mu$ whose scale comes from the geometry of the compact dimensions. The scalar that gets eaten up to 
give the longitudinal polarization of the $Y'$ is a closed string field, and there is no extra higgs particle.
We will leave $M$ as the only free parameter of the model, and use precision electroweak data to constrain its possible values.
In principle, hypercharge can also become massive by mixing with closed string states \cite{GIIQ}, and a topological criterion for this not to happen has been given in \cite{BMMWV}. We assume this property for our model.
  
\par
It is straightforward to identify the photon, and we find that the electric charge is given by
\begin{align}
\frac{1}{e^2}
=\frac {1}{6g_3^2}+\frac{1}{4g_1^2}+\frac{1}{g_2^2}
\end{align}
while $g_3$ and $g_2$ are the strong and weak couplings of the standard model.
From here, it is easy to show that $g_1\sim e/2$ numerically, so the physical angle $\theta_P$ is small. Notice also that the Higgs gets a charge under $Y'$ via this mixing, so that $Y'$ and the $Z$ boson mix. 
After removing the photon, the mass matrix for the neutral vector bosons is governed by \begin{equation}
\begin{pmatrix}\overline{M}^2_Z& \overline{M}_Zg_1 \sin\theta_P v\\
\overline{M}_Zg_1\sin\theta_P v& M^2+g_1^2\sin^2\theta_Pv^2 \end{pmatrix}
\end{equation}
where $4\overline{M}^2_Z= g_2^2 v^2+g_Y^2 v^2$ is the usual tree level formula for the mass of the $Z$ particle in the electroweak theory. It is easy to see that the $Z$ and $Y'$ mixing are suppressed by $\sin\theta_P$ and further suppressed if $M$ is large.
 In practice, the mass we calculate for the $Z$ particle is very close to the standard model. The mass of the $Z$ has been determined experimentally to be $91.1876(21)\;GeV$\cite{PDG}.  At that energy scale, the couplings have values $e=0.31119,\alpha_S=0.1176$, which, constraining to $M_W=80.403\;GeV$\cite{PDG} gives that $g_3=1.215,g_2=0.6596,$ and $g_1=0.1777$.  Plugging these into the mass matrix of the $Z$ and $Z'$ bosons, we see that $M$ can be as low as $800\;GeV$ and still have the mass of the $Z$ be within 1 standard deviation of the experimental value, or as low as $500\;GeV$ to be within 2 standard deviations.   In other words, with $M=500\;GeV$, the $\rho$ parameter is $M_W^2(g_Y^2+g_2^2)/M_Z^2g_2^2=1.00009$, which is within the experimental bound of $1.0000(2)$\cite{PDG}.


\begin{center}\em{Decay Widths}
\end{center}
Switching to the mass basis,
the neutral current part of the covariant derivative is
\begin{align}
\nonumber&\frac{1}{\sqrt{6}}g_3\hat C_\mu Q_{U(3)}+g_2\hat A^3_\mu T^3+g_1\hat B_\mu Q_{U(1)}\end{align}
and it is straightforward to determine the couplings of the various particles to the physical 
$Z$ and the new physical $Z'$ vector boson, as in \cite{IQ,CIK}.

The corrections to the branching ratios and to the decay widths of the $Z$ are very small (of order $0.03\%$),
 well below the experimental bounds of the standard model. Thus we do not get any additional information from this constraint.

The $Z'$ particle coupling to leptons turns out to be very suppressed, while its coupling to quarks is mostly like $U(1)_B$ baryon symmetry and with a strength very close to $g_3/\sqrt 6$. Thus the new $Z'$ particle is leptophobic, and its decay width (as well as  production channels) is dominated by quarks.  Because of this suppression, searches for new particles from $p\overline{p}$ collisions \cite{Abulencia:2005nf} would not be sensitive to this decay.  Since $Z'$ has mixed with $Z$ slightly, there is also some partial width to decay into $W^+W^-$ pairs, however the logitudinal enhancement of the decay from the $W$'s is cancelled by the effective $Z'W$ coupling, and the branching ratio is negligible \cite{delAguila:1986ad}. 

We want to estimate the width of this $Z'$ particle. We will make the approximation that
the width is dominated by decay into quarks, that all quarks are massless (this is reasonable for $M_{Z'}\geq 800 GeV$).
Using standard formulas, we find that approximately $\Gamma_{Z'} \sim 0.15 M_{Z'}$, 
which is about $1/7$ of its mass. This should be contrasted with the $Z$ particle, whose width is about $1/40$ of its mass. The width is small enough that it should be possible to resolve this particle as a slightly wide peak in dijet cross sections at the LHC.

It should also be noticed that the $Z'$ particle of this model couples to all quarks with the same strength, thus it is flavor blind. As such, it does not induce any flavor changing neutral currents (FCNC) and is not ruled out by FCNC bounds. The universality of the $Z'$ couplings to quarks, plus the relation between the decay rates and the standard model couplings make this model very attractive as 
a string inspired extension of the standard model. The relations between the couplings are a consequence of the consistency of string perturbation theory calculations and do not follow from effective field theory arguments alone:
it is easy to show that if we change by hand the coupling of the $U(1)_B$ gauged symmetry in the matter content, it is possible to arrange for mixings with the other $U(1)$ that are sizable and that they would also generically induce large mixings with the $Z$, giving much stronger bounds for $M$, in a way that depends on this extra coupling constant.

\begin{center}
\em Final remarks
\end{center}

The MQSM contains one extra massive $Z'$ particle compared to the standard model. 
Anomaly cancelation by the Green-Schwarz mechanism requires that the effective lagrangian
is non-renormalizable. As such, effective field theory arguments break down at some scale
that can be calculated from the effective lagrangian, and unitarity is restored at this scale by new physics. Our assumption for this paper is that this physics is given by string theory. It is possible that there is some other intermediate renormalizable field theory that describes the MQSM model at this scale and that requires no string theory input. Such a model seems to be very difficult to construct, as the anomalies with the electroweak theory seem to require chiral matter under the Standard Model group. Experimental constraints on such type of matter are very severe. Thus, it is unlikely that this model would be anything other than a string model.

Our model is not free of technical problems. Explaining the smallness of neutrino masses is hard, because the typical dimension five operator that generates their mass would only be suppressed by $M_S$. It is possible that an accidental symmetry prevents a neutrino mass to be generated perturbatively, and then a mass generation mechanism as described recently in \cite{BCW,IU} could provide the required smallness of neutrino masses with an extension of our current model to include another anomalous $U(1)$.

Finally, we would like to point out that it is straightforward to 
make this model supersymmetric. A supersymmetric extension of this model would be interesting, because if the squarks are light enough to be produced in the decay of the $Z'$ particle,  the $Z'$ particle could be used as a factory for superpartners \cite{Kang:2004bz,Baumgart:2006pa}.

D.B. would like to thank M. Cvetic, N.A. Hamed, E. Kiritsis, J. Kumar, R. G. Leigh, F. Quevedo, J. Shelton, D. Stuart,  C. Wagner for discussions related to this work. S.P. would like to thank H. Verlinde and M. Wijnholt for discussions related to this work.  D. B. would also like to thank the Aspen Center for Physics where some of this work was done. Work supported in part by  DOE, under grant DE-FG01-91ER40618

\end{document}